# Modified Design of Microstrip Patch Antenna for WiMAX Communication System

Kumari Nidhi Lal
Wireless communication and Computing
Indian Institute of Information Technology
Allahabad, India
iwc2013017@iiita.ac.in

Ashutosh Kumar Singh
Department of Electronics &
Communication Engineering
Indian Institute of Information Technology
Allahabad, India
ashutosh_singh@iiita.ac.in

*Abstract*—In this paper, a new design for U-shaped microstrip patch antenna is proposed, which can be used in WiMAX communication systems. The aim of this paper is to optimize the performance of microstrip patch antenna. Nowadays, WiMAX communication applications are widely using U-shaped microstrip patch antenna and it has become very popular. Our proposed antenna design uses 4-4.5 GHZ frequency band and it is working at narrowband within this band. RT/DUROID 5880 material is used for creating the substrate of the microstrip antenna. This modified design of the microstrip patch antenna gives high performance in terms of gain and return loss.

## I. INTRODUCTION

Microstrip antenna was proposed in early 1970 [5, 16] and it provides a great revolution in the field of antenna design and research. Nowadays, microstrip patch antenna has become very popular and is widely used in various types of applications[1]. Microstrip antenna provides various features such as comfortability and compatibility. Microstrip patch antennae have many admirable properties such as, high performance, high gain, low cost. The well-known feature of the microstrip patch antennae is that they are reliable and robust in nature. In addition to these properties, it provides an easy user interface and is simple to understand.[3,4]. It has a very simple design and gives very high performance in terms of bandwidth and gain.[12] The topology used for making microstrip patch antenna is U-shaped, which is single narrowband patch antenna.[17]. The E-shaped microstrip patch antenna is used for WLAN applications. [18]While, Dual wideband Stacked patch antenna is used for both WLAN and WIMAX applications[6]. Due to these properties[7], the microstrip patch antenna has become very popular in many applications such as in WiMAX communication system and mobile applications. This antenna is specifically designed for WiMAX communication systems. WiMAX belongs to IEEE 802.16 family of standards. The full form of WiMAX is Worldwide Interoperability for Microwave Access, it provides data rate of 30-40 Mbps, enabling us with interoperable implementations. Microstrip patch antenna has a simple two dimensional geometrical structure. The microstrip antenna is bonded to an insulated dielectric substrate. Rectangular patch is most commonly used by the microstrip antenna. The material, which is used for the patch is copper.

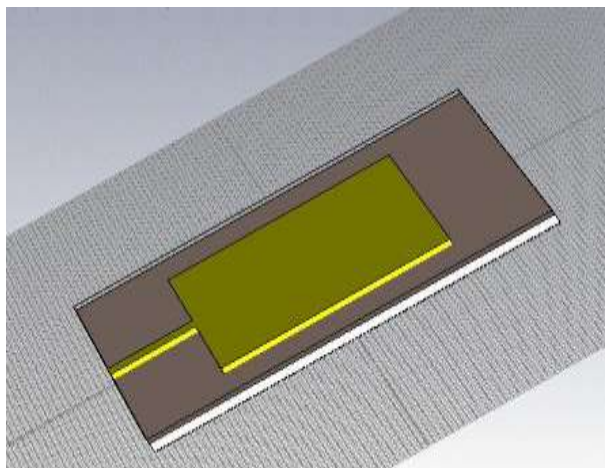
Fig1: a simple microstrip patch antenna

The U-shaped microstrip patch antenna is designed by using substrate material RT/DUROID 5880[6] with permittivity 2.2 and dielectric tangent loss=0.0009.[6]Various iterations have been carried out to calculate the width, length and height of the antenna. The microstrip antenna has a narrowband property, which is a disadvantage of Microstrip patch antenna and shows electromagnetic nature, in some cases.[8,15] There may be many shapes possible for making microstrip patch antenna such as, rectangular[11], circular[8], elliptical, square etc. In this paper, we are designing a Rectangular shaped patch antenna.[2]This U-shaped rectangular microstrip patch antenna is working at a frequency band of[10] 4-4.5 GHz. The height of the substrate is calculated as 2.4 mm, which returns maximum return loss greater than -42dB that improves the gain of the antenna and results high performance.[13,14]The far-field pattern of mirostrip patch antenna also has very good characteristics [25-27].

## II: MODIFIED DESIGN OF U-SHAPED RECTANGULAR MICROSTRIP PATCH ANTENNA

### A. SIMPLE U-SHAPED MICROSTRIP PATCH ANTENNA

[9,13]Simple U-shaped microstrip patch antenna was designed for WiMAX communication system with different dimensions of height and length. The conclusion comes[19-25] out that the antenna gives maximum return loss of -38 dB with height of 2.4mm. The structural design of the simple U-shaped patch antenna is shown in the following figure [27-30].

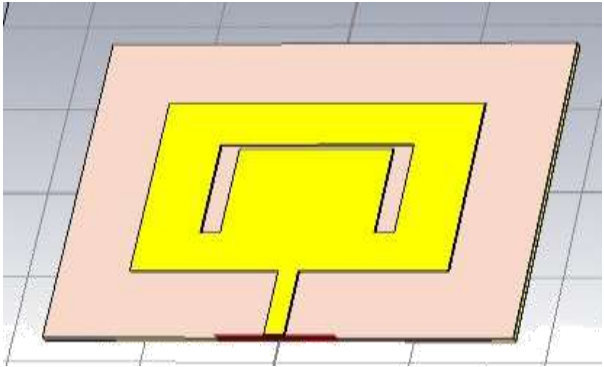
Fig2 : Simple U-shaped microstrip patch antenna

The thickness of the ground plane is about 0.1mm. The height of the substrate is 2.4mm. The material used for making the substrate is RT/DUROID 5880 with permittivity 2.2 and dielectric tangent loss=0.0009.

### B. MODIFIED U-SHAPED D MICROSTRIP PATCH ANTENNA WITH STUBS[9]

In this paper, our aim is to increase the performance of an antenna for using in WiMAX communication application system. The above simple U-shaped patch antenna has maximum return loss of -38dB and maximum gain is about 7.834 dB. In the modified antenna design, all the dimensions are unchanged[30-35] but it provides better performance, better return loss and maximum gain at 4-4.5 GHz frequency band. The structural design of the modified antenna is shown in the following figure.

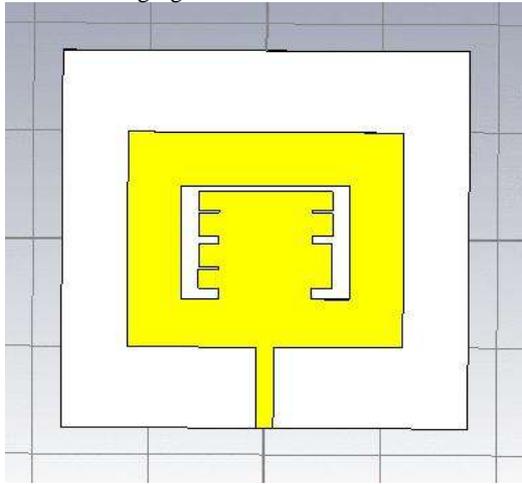
Fig3: Modified design structure of Rectangular microstrip patch antenna

In the modified U -shaped microstrip patch antenna design all the dimensions remain unchanged. It contains seven stubs of different dimensions. It improves the performance of antenna and gives us maximum gain of 8.99dB, which is far better than previous antenna design. It provides us maximum return loss of -43dB which is far better than simple U-shaped microstrip patch antenna[9].

## II. SIMULATION RESULTS AND COMPARISION

### A. RETURN LOSS

In this paper, our aim is to increase the performance of an antenna. In this section, we will compare the return loss for both the simple U-shaped microstrip patch antenna and modified U-shaped microstrip patch antenna.

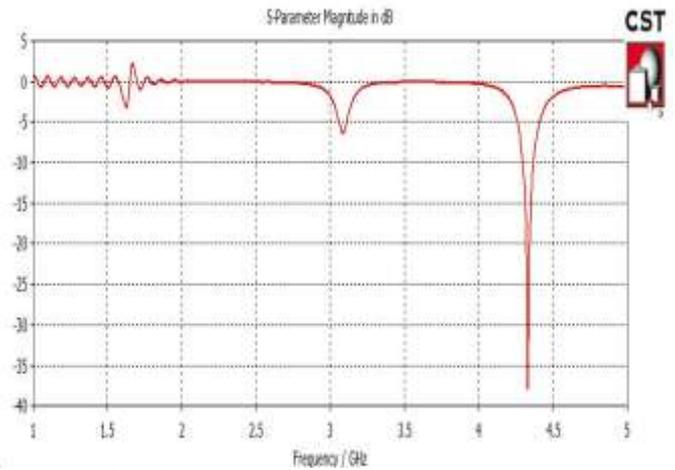
Fig4: Return loss (S-parameter) of simple U-shaped microstrip patch antenna

This is the graphical representation of the return loss(S11) of simple U-shaped microstrip patch antenna, which has a maximum value of -38dB.

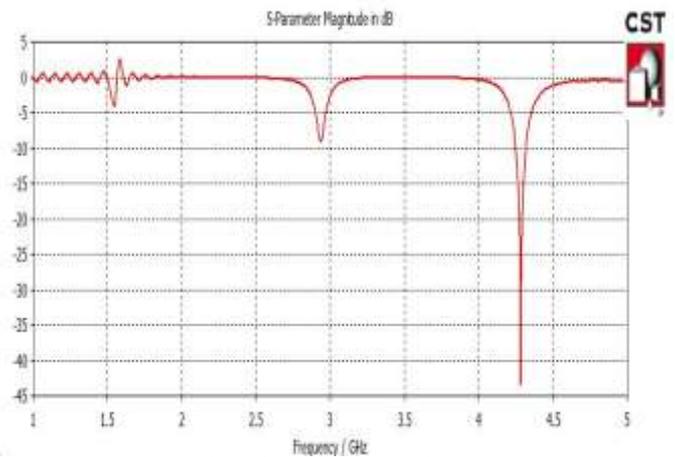
Fig5: Return loss of modified microstrip patch antenna

The above graph represents the return loss of modified U-shaped microstrip patch antenna, which have seven stubs with different dimensions. It provides us the maximum return loss of -43dB which is better than simple U-shaped microstrip patch antenna.

### A. SIMULATION OF GAIN IN DECIBELS AND PLOTTED FAR -FIELD (RADIATION PATTERN)

The gain of an antenna is described as the intensity of radiation of the antenna in a particular direction, which relates the concept of directivity and electrical efficiency of antenna. Radiation pattern is the representation of obtained gain, including direction as a function. The maximum gain obtained from the simple U-shaped microstrip patch antenna is about 7.83 dB. While, the maximum gain obtained from the modified design of U-shaped microstrip patch antenna is about 8.99dB, so hence the modified design gives better performance than the simple design. Hence it is more advisable for the WiMAX communication system and other useful applications.

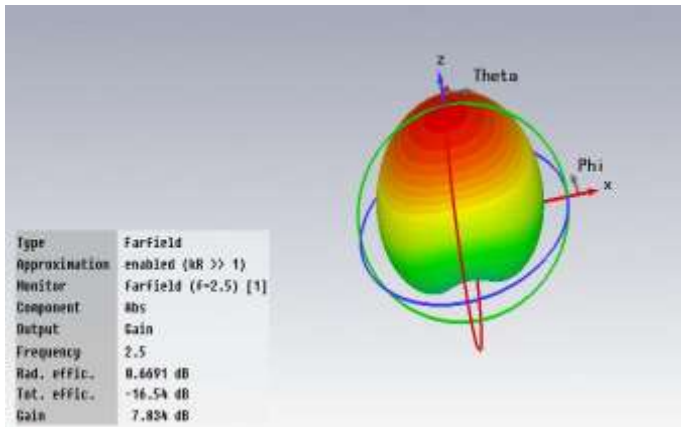

Fig 6: simple U-shaped microstrip patch antenna far -field pattern and gain

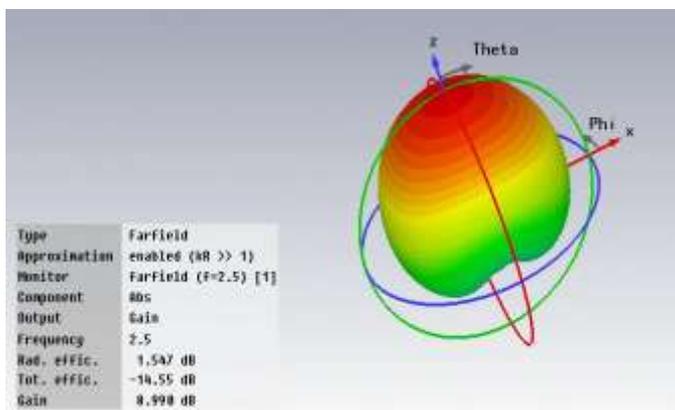

Fig 7: modified U-shaped Rectangular microstrip patch antenna's far field pattern and gain

The performance of the WiMAX applications depend on certain parameters such as gain, radiation pattern, return loss. These parameters depend on the simulation performance and design techniques of the microstrip patch antenna. The gain is the one of the most important property of the microstrip patch antenna. In the above figure, the received gain is about 8.99dB, which is far better than the smiple microstrip patch antennna. The above figure shows the simulation of gain in theta. In simulation results, the gain of modified U-shaped microstrip patch antenna is greater than simple U-shaped microstrip patch antenna. Therefore this design of antenna gives us better results and performance for use in the WiMAX applications.

### D. PARAMETER SPECIFICATION AND VALUES USED IN THE DESINGN OF ANTENNA

List of parameters used in the design of simple U-shaped microstrip patch antenna are given below in table C.1

| Frequency | 4-4.5GHZ |
|---|---|
| Substrate material | RT/duro 5880 |
| Height of the substrate | 2.4mm |
| Patch width | 47.43mm |
| Patch height | 39.098mm |
| Thickness of the ground | 0.1mm |
| Material of the patch | copper |
| Feed line(l3,w3) | 28.1,3mm |
| L1(l1,w1) | 20,3mm |
| L2(l2,w2) | 20,3mm |
| W2(l,w) | -29,-0.7mm |

List of Parameters used in the design of modified U-shaped microstrip patch antenna are given below in table C.2

| Frequency | 4-4.5GHZ |
|---|---|
| Substrate material | RT/duro 5880 |
| Height of the substrate | 2.4mm |
| S1(l,w) | 4,0.5mm |
| S2(lw) | 4,1.4mm |
| S3(lw) | 4,0.5mm |
| S4(l,w) | 4,2mm |
| S5(l,w) | 4,0.5mm |
| S6(l,w) | 4,1.4mm |
| S7(l,w) | 4,2mm |

The S1, S2, S3, S4, S5, S6, S7 are the stubs in the modified U-shaped patch antenna starting from the left one cut and so on.

### III. CONCLUSION

The microstrip patch antenna is most widely used antenna design in the WiMAX communication system. In this paper,we have increased the performance of the antenna by improving certain essential parameters of the antenna such as radiation pattern, gain and return loss. We do some modifications here without changing the dimensions of the microstrip patch antenna.This great performance is achieved by modifying the structure of the patch antenna. The improved design give us gain of 8.99dB with maximum return loss -43dB, which greatly helps to enhance the performance of the antenna so that it can be efficiently used in the WiMAX communication system and other applications.